\documentclass[twocolumn]{aastex7}

\usepackage{latexsym}
\usepackage{amsmath}
\usepackage{graphicx}
\usepackage{longtable,booktabs}
\usepackage{natbib}
\usepackage{CJK}

\begin{document}

\title{Identifying Merger-Driven and Collapsar-Driven Gamma-Ray Bursts with Precursor based Solely on Prompt Emission}

\tighten

\correspondingauthor{Pak-Hin Thomas Tam}
\email{tanbxuan@mail.sysu.edu.cn}

\author[0009-0007-0686-3906,sname='Zhu',gname='Si-Yuan']{Si-Yuan Zhu}
\affiliation{School of Physics and Astronomy, Sun Yat-Sen University, Zhuhai 519082, China}
\affiliation{CSST Science Center for the Guangdong-Hong Kong-Macau Greater Bay Area, Sun Yat-sen University, Zhuhai 519082, China}
\email{zhusy37@mail2.sysu.edu.cn}

\author[0000-0002-1262-7375]{Pak-Hin Thomas Tam}
\affiliation{School of Physics and Astronomy, Sun Yat-Sen University, Zhuhai 519082, China}
\affiliation{CSST Science Center for the Guangdong-Hong Kong-Macau Greater Bay Area, Sun Yat-sen University, Zhuhai 519082, China}
\email{tanbxuan@mail.sysu.edu.cn}

\author[0000-0002-5936-8921]{Fu-Wen Zhang}
\affiliation{College of Physics and Electronic Information Engineering, Guilin University of Technology, Guilin 541004, China}
\affiliation{Key Laboratory of Low-dimensional Structural Physics and Application, Education Department of Guangxi Zhuang Autonomous Region, Guilin 541004, China}
\email{fwzhang@pmo.ac.cn}

\author[0009-0003-8397-4125]{Hui-Ying Deng}
\affiliation{School of Physics and Astronomy, Sun Yat-Sen University, Zhuhai 519082, China}
\affiliation{CSST Science Center for the Guangdong-Hong Kong-Macau Greater Bay Area, Sun Yat-sen University, Zhuhai 519082, China}
\email{denghy59@mail2.sysu.edu.cn}

\author[0000-0002-9725-2524]{Bing Zhang}
\affiliation{The Hong Kong Institute for Astronomy and Astrophysics, The University of Hong Kong, Pokfulam Road, Hong Kong, China}
\affiliation{Department of Physics, the University of Hong Kong, Pokfulam Road, Hong Kong, China}
\email{bzhang1@hku.hk}

\begin{abstract}
Gamma-ray bursts (GRBs) are generally classified as Type~I GRBs, which originate from compact binary mergers, and Type~II GRBs, which originate from massive collapsars.
The traditional correspondence between short--Type~I GRBs and long--Type~II GRBs, separated by a duration of 2 seconds, has been challenged by recent observations of long GRBs associated with kilonovae (i.e., Type~I-L GRBs) and a short GRB associated with a supernova.
In this paper, we focus on GRBs with precursor emission (PE) and compile 366 GRBs detected by Fermi/GBM.
Applying the unsupervised machine learning methods t-SNE and UMAP, we are able to distinguish Type~I (including subclass Type~I-L) and Type~II GRBs for the first time and identify PE as a key feature for distinguishing GRBs of different origins.
Inspired by results of machine learning, we propose a diagnostic parameter, the $E_{\rm p,ME}$-precursor index ($EPI$), defined as ${\rm log_{10}}(E_{\rm p,ME}^{2}/(T_{\rm 100,PE}T_{\rm 100,QE1}^{1/2}T_{\rm MVT,PE}))$, where most Type~I GRBs have $EPI > 6.2$ and most Type~II GRBs have $EPI < 6.2$. This parameter can help the community to diagnose the origin of any GRB with PE based solely on its prompt emission and rapidly plan for follow-up observations.
The validation using Swift GRBs provides illustrative evidence that our method may also be applicable to GRBs observed by instruments other than Fermi.

\end{abstract}


\section{Introduction} \label{sec:introduction}
Gamma-ray bursts (GRBs) are generally believed to have two origins \citep{2006Natur.444.1010Z,2009ApJ...703.1696Z}: compact binary mergers \citep[Type~I GRBs;][]{1986ApJ...308L..43P,1992ApJ...395L..83N} and massive collapsars \citep[Type~II GRBs;][]{1993ApJ...405..273W,1999ApJ...524..262M}.
Type~I GRBs are associated with gravitational waves (GWs)/kilonovae \citep[KNe;][]{2017PhRvL.119p1101A,2017ApJ...851L..18W}, while Type~II GRBs are associated with supernovae \citep[SNe;][]{1998Natur.395..670G,2006ARA&A..44..507W}.

Since the BATSE era, it has become a paradigm that GRBs are classified based on their duration ($T_{90}$) into long GRBs (LGRBs; $T_{90} > 2$ s) and short GRBs \citep[SGRBs; $T_{90} < 2$ s;][]{1993ApJ...413L.101K}.
Type~I and Type~II GRBs exhibit similar physical properties to the majority of SGRBs and LGRBs, respectively.
Therefore, it is generally believed that the vast majority of SGRBs originate from mergers (Type~I-S GRBs), whereas most LGRBs originate from collapsars (Type~II-L GRBs).

However, recent observations revealed that some LGRBs, including GRB~060614, GRB~211211A and GRB~230307A, are associated with KNe \citep[Type~I-L GRBs;][]{2015NatCo...6.7323Y,2022Natur.612..223R,2022Natur.612..232Y,2024Natur.626..742Y,2024Natur.626..737L}, while a SGRB, GRB~200826A, is associated with SN \citep[Type~II-S GRBs;][]{2021NatAs...5..917A,2021NatAs...5..911Z}.
These events challenge the traditional correspondence between short-Type~I and long-Type~II, suggesting that the $T_{90}$ classification alone cannot distinguish GRBs of different physical origins.
In addition, the whole emission (WE) of some GRBs is composed of precursor emission (PE), main emission (ME), and extended emission (EE), which causes $T_{90}$ to be an unreliable measure of their true duration.

Additionally, some differences between SGRBs and LGRBs are used to distinguish GRBs with different origins.
Compared to LGRBs, SGRBs generally exhibit negligible or negative spectral lags \citep[$\tau$;][]{2006MNRAS.367.1751Y}, shorter minimum variability timescales \citep[MVT;][]{2013MNRAS.432..857M}, harder spectra \citep{2012ApJ...750...88Z}, and distinct correlations between the isotropic energy and the peak energy in the rest frame \citep[$E_{p,z}$--$E_{\rm iso}$ correlation;][]{2002A&A...390...81A,2013ApJ...763...15Q,2023ApJ...950...30Z}.
However, some Type~II GRBs also show negative $\tau$ \citep{2023MNRAS.524.1096L} and relatively short MVTs \citep{2023ApJ...954L...5V}.
Furthermore, some classification methods based solely on spectral hardness and the $E_{p,z}$--$E_{\rm iso}$ correlation also fail to simultaneously distinguish Type~I GRBs from Type~II GRBs \citep{2010ApJ...725.1965L,2020MNRAS.492.1919M,2024ApJ...976...62Z}.

The t-distributed stochastic neighbor embedding \citep[t-SNE;][]{2008JMLR.9.2579M,2014JMLR.15.3221M} and the Uniform Manifold Approximation and Projection \citep[UMAP;][]{2018arXiv180203426M} are two powerful unsupervised machine learning algorithms.
They have been widely used in GRB classification owing to their effectiveness in mapping neighboring data from high-dimensional space onto a two-dimensional representation \citep{2020ApJ...896L..20J,2023ApJ...949L..22D,2023ApJ...945...67S,2024ApJ...974...55D,2024MNRAS.532.1434Z,2025A&A...702A.173Z}.
In addition, \citet{2023ApJ...959...44L} employed supervised machine learning to classify GRBs, and their feature importance analysis indicated that the prompt emission is a key discriminator between Type~I and Type~II GRBs.
Recently, \citet{2025MNRAS.541.3236Z} identified Type~I-L GRBs from three-episode GRBs based solely on prompt emission and suggested that PE may play a key role in distinguishing Type~I-L and Type~II-L GRBs.
However, the sample they analyzed was relatively small and did not include Type~I-S GRBs, limiting their ability to provide a comprehensive picture of GRBs with different origins.

In this paper, we focus on GRBs with PE to investigate whether merger and collapsar progenitors produce distinct prompt emission.
We further propose a new diagnostic method that can distinguish between Type~I and Type~II GRBs solely based on the properties of the PE and ME.
We suggest that PE may serve as a newly recognized observational indicator, enabling rapid, yet secure, identification of the origin of a GRB, thereby providing strong guidance for subsequent multi-wavelength follow-up observations.
Moreover, our results provide evidence that the PE of Type~I GRBs have different physical origins from ME, potentially arising from interactions in the pre-merger binary system.

The structure of this paper is organized as follows.
In Section~\ref{sec:data}, we describe the sample selection criteria and the data analysis methods.
In Section~\ref{sec:ml-classification}, we present the classification results based on machine learning and describe the origins and properties of different types of GRBs.
In Section~\ref{sec:tr-classification}, we propose a classification scheme that is based on prompt emission only, enabling a rapid distinction between GRBs of different origins.
The discussion is presented in Section~\ref{sec:discussions}, and the conclusions are presented in Section~\ref{sec:conclusions}.

\section{Data Analysis and Methodology} \label{sec:data}
The Fermi satellite, launched in 2008, has collected an extensive dataset of GRB observations with spectral coverage ranging from 8 keV--300 GeV.
In this work, we used the time-tagged event (TTE) data of Gamma-ray Burst Monitor (GBM) for temporal and spectral analyses with an energy range of 8 keV--40 MeV \citep{2009ApJ...702..791M}.

\subsection{Light Curve Extraction and Temporal Analysis}\label{subsec:lightcurve}
We comprehensively searched for GRBs with PE in the Fermi/GBM catalog from August 2008 to April 2025 \citep{2020ApJ...893...46V}.
The light curves in the 8--1000 keV energy range were extracted using the standard fermi tool for Python \texttt{GBM Data Tools}.
We employed the Bayesian blocks algorithm to determine the optimal change points of light curves \citep{2013ApJ...764..167S}, whose time resolutions are listed in Table \ref{table:properties}.
Based on the resulting segmentation, we divided the WE into several distinct episodes: PE, ME, and EE, as well as the quiescent episodes (QEs) separating the PE and ME, and the ME and EE.
Examples of such GRBs are shown in Figure~\ref{figure:lightcurve}.

The identification of PE relies exclusively on light curves.
The PE is defined as a distinct emission episode preceding the ME, separated by a QE, and showing a distinct pulse structure.
To minimize random coincidences caused by statistical fluctuations, we require the duration of the QE to be at least eight times the time resolution, and the QE must be identified as a single Bayesian block, indicating that no statistically significant flux variation is detected during the interval (i.e., stable emission).
In addition, \citet{2025ApJ...979...73W} found that weak emission with signal-to-noise ratio (SNR) of $\sim 3$ can still be present during the QE.
 Therefore, instead of requiring the emission to drop strictly to the background level, we empirically define the QE as an interval exhibiting relatively stable emission close to the background, i.e., SNR below around $3\sigma$ \citep{2025ApJ...979...73W,2025MNRAS.541.3236Z}.
To further ensure the independence of the PE, we require that when the QE is consistent with the background level, the peak count of the PE may be comparable to that of the ME, though with lower flux. 
Otherwise, the peak count of the ME must be significantly higher than that of the PE, and the PE peak must exceed the background by at least $3\sigma$.

Although our primary goal was to search for GRBs with PE, some of these GRBs also exhibit independent pulses or long, weak tail following the ME (referred to as EE). 
The criterion for identifying EE is that it must occur after the ME and include a QE, without imposing a specific SNR threshold.
Since the initial screening relies on manual inspection, some cases may have been missed.
GRBs with low SNR, extremely complex backgrounds, or incomplete data were also excluded.
In total, 366 GRBs with PEs were identified from 3982 GRBs, including 29 GRBs from \citet{2025MNRAS.541.3236Z}.
Additionally, GRB~060614, a confirmed merger-driven LGRB with a PE, was also included in the sample.
The final sample thus consists of 367 GRBs, including three LGRBs associated with KNe (GRBs-KNe) and six LGRBs associated with SNe (GRBs-SNe).

We analyze the light curves of the PE, ME, EE, and QE.
The duration of each episode is defined as the time difference between the start and end times.
We calculated the durations of the PE ($T_{\rm 100,PE}$), ME ($T_{\rm 100,ME}$), WE ($T_{\rm 100,WE}$), and the QE between the PE and ME ($T_{\rm 100,QE1}$).
The spectral lag is defined as the time delay of low-energy photons with respect to high-energy photons \citep{1986ApJ...301..213N,1995A&A...300..746C}.
We used the cross-correlation function (CCF) to calculate $\tau$ of the 50--100 keV relative to the 25--50 keV ($\tau_{32}$).
For GRB~060614, $\tau_{32}$ was calculated using the same energy channels.
Note that $\tau$ was not computed if the CCF cannot be reliably fitted at a time resolution below 0.256 s.

The MVT represents the shortest resolvable variability timescale, typically corresponding to the rise time of the shortest pulse \citep{2014ApJ...787...90G,2015ApJ...811...93G}.
We employed Bayesian block method to identify the shortest pulse in the PE and ME, and we denote their rise times as $T_{\rm MVT,PE}$ and $T_{\rm MVT,ME}$, respectively, \citep{2023ApJS..268....5X,2025ApJ...979...73W}.
Our results are consistent with the MVT values of Fermi/GBM GRBs reported by \citet{2025A&A...702A..95M}, despite being derived using a different method.

\subsection{Spectral Analysis and Properties}\label{subsec:rest}
We fitted the spectra of the ME for each GRB using the \texttt{PyXspec} based on the standard \texttt{HEASOFT}.
We used 10--900 keV for the NaI detectors \citep[excluding 30--40 keV to avoid the iodine K edge;][]{2009ApJ...702..791M} and 0.3--40 MeV for the BGO detectors.

Generally, the spectra of GRBs are non-thermal, with quasi-thermal components also detected in some bursts.
Therefore, we employed five spectral models, including two non-thermal models: Band \citep{1997ApJ...486..928B} and cutoff power-law (CPL), as well as three combined models that incorporate both non-thermal and quasi-thermal blackbody (BB) components: Band+BB, CPL+BB, and PL+BB.
The best-fit model was selected based on the minimum Bayesian Information Criterion (BIC).
If certain parameters in the model cannot be constrained, then this model will be excluded.
The temporal and spectral analysis results are list in Table~\ref{table:properties}.

\begin{table*}
	\caption{The temporal and spectral properties of GRBs}
	\centering
	\label{table:properties}
	\scriptsize
	\setlength{\tabcolsep}{4.5pt}
	\renewcommand{\arraystretch}{1.2}
	\begin{tabular}{llcccccccccccc}
		\hline
		GRB & Fermi~ID       & $T_{\rm res}$ & $T_{\rm 100,PE}$ & $T_{\rm 100,ME}$ & $T_{\rm 100,WE}$ & $T_{\rm 100,QE1}$ & $T_{\rm MVT,PE}$ & $T_{\rm MVT,ME}$ & $E_{\rm p,ME}$ & $\tau_{\rm 32,ME}$ & $\tau_{\rm 32,WE}$ & $T_{\rm 90}$ & Type\\
		& & (s) & (s) & (s) & (s) & (s) & (s) & (s) & (keV) & (s) & (s) & (s) & \\
		\hline
		060614A & \nodata & 0.016  & 0.128     & 6.08      & 124.16    & 0.544     & 0.072     & 0.008     & 302       & 34.51     & 35.09     & 123.65 & I-L\\
		080715A& 080715950  & 0.032  & 3.84      & 4.128     & 108.896   & 2.688     & 1.936     & 0.112     & 273.56    & 28.72     & 34.67     & 7.87 & II\\
		080830A& 080830368  & 0.064  & 5.056     & 9.664     & 49.28     & 3.584     & 2.016     & 2.272     & 272.95    & -74.83    & -240.13   & 40.9 & II \\
		080913B& 080913735  & 0.016  & 8.608     & 16.528    & 28        & 2.864     & 0.296     & 0.424     & 102.94    & 49.27     & 46.84     & 41.22 & II\\
		081216A& 081216531  & 0.008  & 0.152     & 0.464     & 1.096     & 0.48      & 0.012     & 0.116     & 1179.19   & 19.05     & 19.34     & 0.77 & I-S\\
		\nodata\\
		\hline
		\multicolumn{12}{l}{This table is published in its entirety in the machine-readable format. A full machine-readable version is available online.}\\
	\end{tabular}
\end{table*}

\subsection{Machine Learning}\label{subsec:ml}
To uncover potential structures in the data, we employed t-SNE and UMAP for dimensionality reduction and visualization of the high-dimensional GRB parameters. 
The two methods provide a cross-check and help minimize possible artifacts.
Note that the coordinates obtained from t-SNE and UMAP have no physical meanings.
More details on the t-SNE and UMAP methods can be found in \cite{2018arXiv180203426M} and references therein.
Prior to the analysis, the dataset is logarithmically transformed (except for $\tau_{\rm 32,ME}$ and $\tau_{\rm 32,WE}$, which include negative values) and standardized using the Z-score method.

The Python module $scikit$-$learn$ and $umap$-$learn$ were used for t-SNE and UMAP analyses, respectively.
The $perplexity$ is the primary hyperparameter controlling the embedding results in t-SNE, while $n\_neighbors$ and $min\_dist$ play similar roles in UMAP.
In this work, $min\_dist$ was fixed at 10$^{-4}$.

We employed the hierarchical density-based spatial clustering of applications with noise \citep[HDBSCAN;][]{2017JOSS....2..205M} algorithm to classify clusters in the embedded spaces of t-SNE and UMAP.
HDBSCAN is an unsupervised clustering algorithm that can automatically determine the optimal number of clusters in the data and assign a cluster label to each GRB.
The HDBSCAN analysis was performed using the $scikit$-$learn$.

\section{GRB Classification based on Machine Learning} \label{sec:ml-classification}

We employed t-SNE ($perplexity = 6$) and UMAP ($n\_neighbors=6$) to embed the 9 parameters of GRBs ($T_{\rm 100,PE}$, $T_{\rm 100,ME}$, $T_{\rm 100,WE}$, $T_{\rm 100,QE1}$, $T_{\rm MVT,PE}$, $T_{\rm MVT,ME}$, $\tau_{\rm 32,ME}$, $\tau_{\rm 32,WE}$, and $E_{\rm p,ME}$) into a two-dimensional space for visualization, respectively.

\begin{figure*}
	\centering
	\includegraphics[angle=0,scale=0.72]{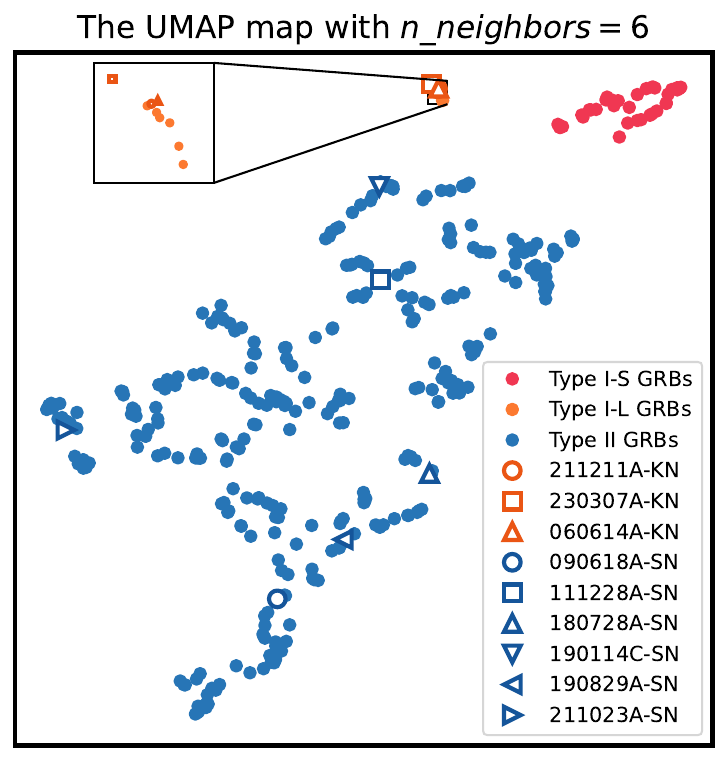}
	\includegraphics[angle=0,scale=0.72]{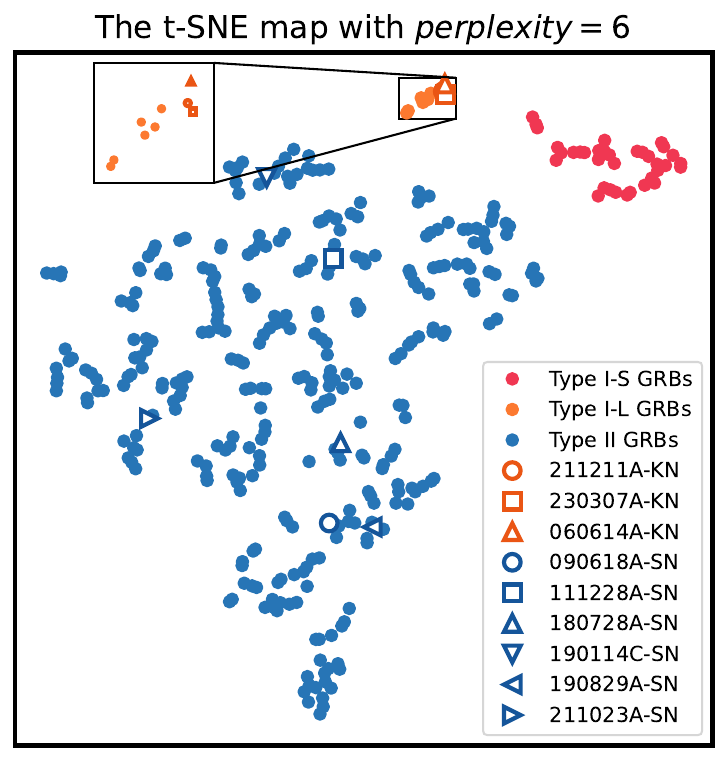}
	\caption{The UMAP and t-SNE maps based on 9 parameters. The red, orange, and blue markers represent the three clusters classified by HDBSCAN. The hollow orange and blue markers show observationally confirmed Type~I-L GRBs-KN and Type~II GRBs-SN.}
	\label{figure:classification}
\end{figure*}

As shown in Figure~\ref{figure:classification}, applying the HDBSCAN algorithm to the t-SNE and UMAP maps reveals three distinct clusters (or classes) of GRBs. First of all, the three observationally-established long-duration GRBs-KNe and the six GRBs-SNe are classified into two distinct clusters, in orange and blue respectively, suggesting that these two clusters indeed correspond to different classes of progenitors. 
We refer to other members in the two clusters as Type~I-L GRB and Type~II-L GRB candidates, respectively.
Although no GRBs with confirmed origins are found in the third (red) cluster, it includes three SGRBs with well-observed host galaxies--GRB~090510, GRB~111117A, and GRB~161001A--which are generally considered to be Type~I-S GRBs.
Their host galaxies are all old galaxies with low specific star formation rates (sSFRs) and large offsets from the galactic centers, features that strongly suggest an origin from compact binary mergers \citep{2018A&A...616A..48S,2020ApJ...897..154L,2022ApJ...940...56F,2022ApJ...940...57N}. 
Therefore, we refer to this cluster as Type~I-S GRB candidates.
Among the 367 GRBs, the classification based on the machine learning identifies 35 Type~I-S GRBs, 9 Type~I-L GRBs, and 323 Type~II GRBs.

To study the properties of these three classes of GRBs, we depict a comprehensive statistical plot of the 9 prompt emission parameters, as shown in Figure~\ref{figure:par}. 
Evidently, Type~I-S GRBs exhibit significantly shorter $T_{\rm 100,PE}$, $T_{\rm 100,ME}$, $T_{\rm 100,WE}$, $T_{\rm 100,QE1}$, $T_{\rm MVT,PE}$, and $T_{\rm MVT,ME}$ compared to Type~II GRBs. 
Their $\tau_{\rm 32,ME}$ and $\tau_{\rm 32,WE}$ values are distributed around zero, and their $E_{\rm p,ME}$ values are generally higher than those of Type~II GRBs. 
These properties are consistent with the conventional distinctions between short and long GRBs.
Type~I-L GRBs, except for $T_{\rm 100,ME}$ and $T_{\rm 100,WE}$ which share similar distributions with Type~II GRBs, exhibit nearly identical properties to Type~I-S GRBs. This is consistent with the traditional understanding of Type~I-L GRBs.
This similarity indicates that Type~I-S and Type~I-L GRBs likely arise from the same merger-driven origin. Type~I-L GRBs may represent a long-duration subclass that is not powered by accretion \citep{2025JHEAp..45..325Z}.

\begin{figure*}[htbp!]
	\centering
	\includegraphics[angle=0,scale=0.72]{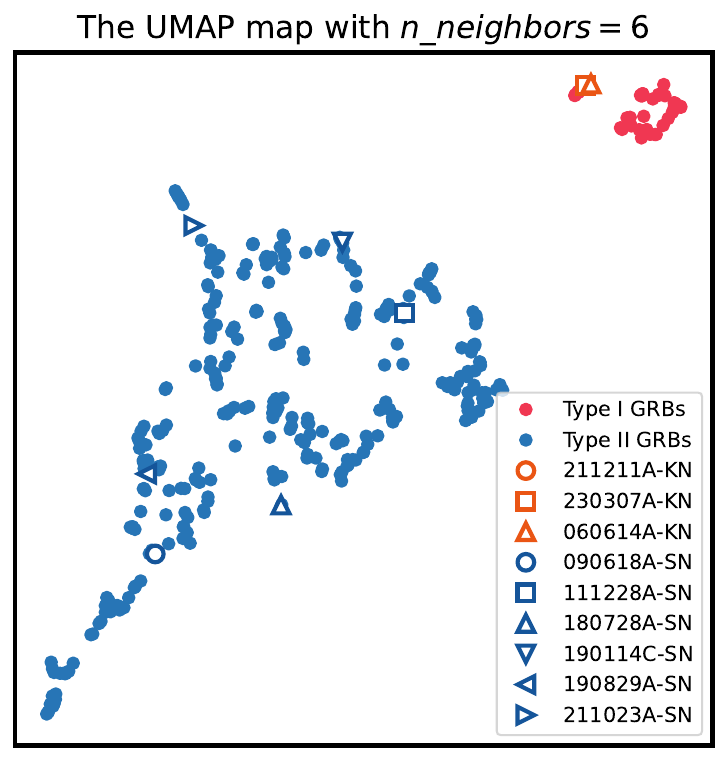}
	\includegraphics[angle=0,scale=0.72]{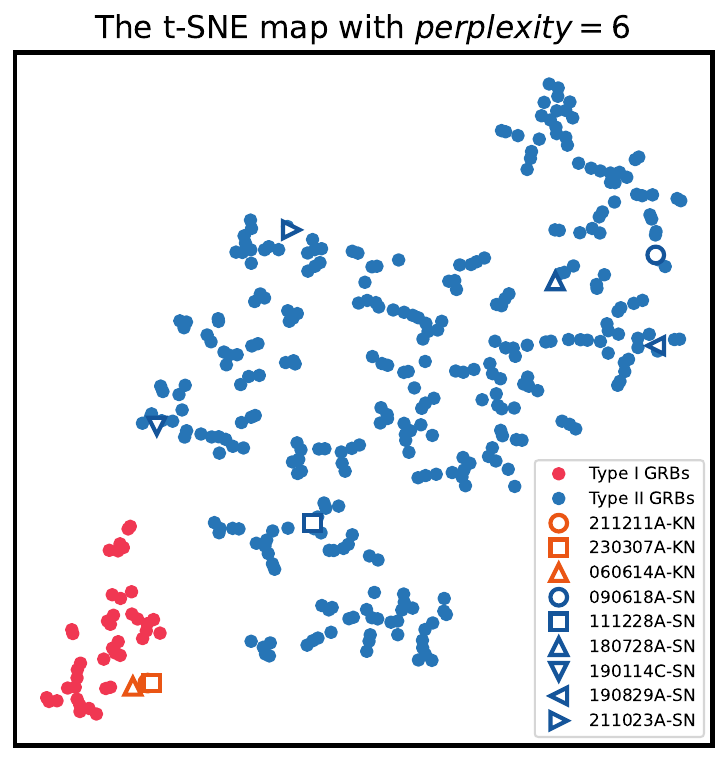}
	\caption{The UMAP and t-SNE maps based on 8 parameters beside $T_{\rm 100,WE}$. The red and blue markers represent the two clusters classified by HDBSCAN.}
	\label{figure:classification8}
\end{figure*}

\citet{2025MNRAS.541.3236Z} identified six Type~I-L GRB candidates.
Among them, GRB~170228A, GRB~200311A, GRB~200914A, and GRB~211019A are also classified as Type~I-L GRBs, whereas GRB~090831 and GRB~180605A, characterized by short PE, ME, and QE durations and a long EE, are re-classified as Type~I-S GRBs in this work.
The difference arises because \citet{2025MNRAS.541.3236Z} considered only three-episode GRBs, so most Type~I-S GRBs without EE were excluded from their sample, leading to the misclassification of GRB~090831 and GRB~180605A.
In addition, some Type~I-S GRBs also exhibit relatively long-duration ME or WE, but not both. 
This suggests that Type~I-L represents a subclass of Type~I GRBs characterized by both a longer $T_{\rm 100,ME}$ and a longer $T_{\rm 100,WE}$.
Therefore, we suggest an empirical classification for those Type~I GRBs with $T_{\rm 100,ME} > 2$ s and $T_{\rm 100,WE} > 10$ s as Type~I-L GRBs, while the remaining are classified as Type~I-S GRBs (see Appendix \ref{section:sl} for more details).

Additionally, GRB~131215B and GRB~170412B are identified as two new Type~I-L GRB candidates. 
GRB~131215B consists of a 0.16 s PE, a 28.51 s ME, resulting in a total duration of 31.2 s, with $T_{\rm 100,QE1} = 2.53$ s. 
GRB~170412B consists of a 0.34 s PE, a 33.31 s ME, resulting in a total duration of 35.47 s, with $T_{\rm 100,QE1} = 1.82$ s. 
Their light curves exhibit short PEs and QEs but much longer MEs, consistent with the typical properties of Type~I-L GRBs.
Meanwhile, their short MVTs ($T_{\rm MVT,PE}$ are 0.11 and 0.17 s; $T_{\rm MVT,ME}$ are 0.08 s and 0.1 s) and the small spectral lags of the ME ($\tau_{\rm 32,ME}$ are 52.6 ms and -2.59 ms) further support this classification.
However, since their EEs cannot be clearly distinguished (possibly blended with the ME), they were not included in \citet{2025MNRAS.541.3236Z}.
We note that GRB~170228A, GRB~200311A and GRB~200914A have been independently identified as TypeI-L GRBs and show prompt emission properties similar to those of Type~I GRBs in several aspects, such as the $E_{\rm p,z}$--$E_{\rm iso}$ relation and the power spectral density \citep{2025ApJ...993...89T,2025MNRAS.541.3236Z}; these results further support the reliability of the classification scheme presented in this section.

Although no distinct Type~II-S GRB population emerges from our results, we searched for potential Type~II-S GRB candidates. 
Based on the threshold inferred from GRB~200826A with $T_{\rm 100,ME} \sim 2.5$ s, we suggested 15 Type~II-S candidates.
These bursts show temporal properties similar to Type~I-S GRBs but display significantly softer spectra than typical Type~I GRBs.
Among them, GRB~180703B shares properties with GRB~200826A and has also been suggested as a Type~II-S candidate by \citet{2024ApJ...976...62Z}.
These candidates lie near the boundary of the Type~II GRB population, suggesting that their differences from normal Type~II GRBs are modest compared with the contrast between Type~I-L and Type~I-S GRBs. 
Since GRB~200826A lacks a PE, confirming the existence of a genuine Type~II-S population remains challenging.

We further employed t-SNE and UMAP to embed 8 GRB parameters excluding $T_{\rm 100,WE}$, HDBSCAN separates the GRBs into two clusters.
As shown in Figure~\ref{figure:classification8}, Type~I-S and Type~I-L GRBs form a single cluster, whereas Type~II GRBs remain distinct, effectively distinguishing merger-driven and collapsar-driven GRBs.

We find that Type~I GRB are primarily distinguished from Type~II GRBs by shorter $T_{\rm 100,PE}$, $T_{\rm 100,QE1}$, and $T_{\rm MVT,PE}$.
It is worth noting that one exceptional Type~I-S GRB, GRB~131209B, shows a relatively long $T_{\rm 100,QE1}$ of 15.42 s, while its $T_{\rm 100,PE}$ and $T_{\rm MVT,PE}$ are only 0.19 s and 0.1 s, respectively.
In addition, we confirm that when $T_{\rm 100,PE}$, $T_{\rm 100,QE1}$, and $T_{\rm MVT,PE}$ are excluded, neither the t-SNE nor the UMAP algorithm can reproduce the aforementioned classification results.
These results provide new insight that the PE of merger-driven GRBs may originate from different physical processes compared to those of collapsar-driven GRBs, which suggests that PE can serve as a potential diagnostic indicator for distinguishing between merger and collapsar origins.

Since machine learning results are inevitably affected by hyperparameters and random initialization, we tested the stability of our clustering by changing key hyperparameters (e.g., setting $perplexity$ and $n\_neighbors$ to 5, 6, and 7, respectively) and using different random initializations. 
The results show that t-SNE and UMAP produce consistent classification results across different hyperparameter settings.

Potential instrument-related biases may also influence our results.
GRB~060614 is the only burst in our sample that was detected by Swift/BAT rather than Fermi, and its PE properties are derived from the 15--350 keV energy band.
Removing this burst does not change the classifications, indicating that our results are not sensitive to its inclusion. 
While this suggests potential applicability to other instruments, caution is still required when extending the analysis.

Potential redshift biases may diminish the differences in the parameters used for machine learning between the two types (e.g., Type II GRBs occur at systematically higher redshifts than Type I GRBs), and therefore may affect the classification.
We performed redshift corrections for the 30 GRBs with known redshifts and repeated the machine learning analysis in the rest frame.
The results still reveal two clearly separated clusters corresponding to Type I and Type II GRBs.
The classifications remain unchanged for almost all bursts, except for GRB 230818A, which shifts from the Type II to the Type I cluster after the correction. 
Analysis in the observer frame shows a consistent clustering pattern, suggesting that this change is more likely caused by the limited size of the redshift-known sample rather than by redshift bias.
Overall, this indicates that redshift effects do not significantly affect the robustness of our classification.

\section{A New Rapid Classification Scheme} \label{sec:tr-classification}

The machine learning results indicate that no single parameter can clearly separate GRBs of different origins, and multiple parameters are therefore required for classification.
As mentioned in the previous section, PE properties provide key discriminators, with Type~I GRBs exhibiting shorter $T_{\rm 100,PE}$, $T_{\rm 100,QE1}$, and $T_{\rm MVT,PE}$.
To quantify these features, we define a precursor index,
\begin{equation}
    PI = \left( \frac{T_{\rm 100,PE}}{\rm s} \right) \times \left( \frac{T_{\rm 100,QE1}}{\rm s}\right)^{1/2} \times \left( \frac{T_{\rm MVT,PE}}{\rm s}\right).
\end{equation}

\begin{figure*}
	\centering
	\includegraphics[angle=0,scale=0.43]{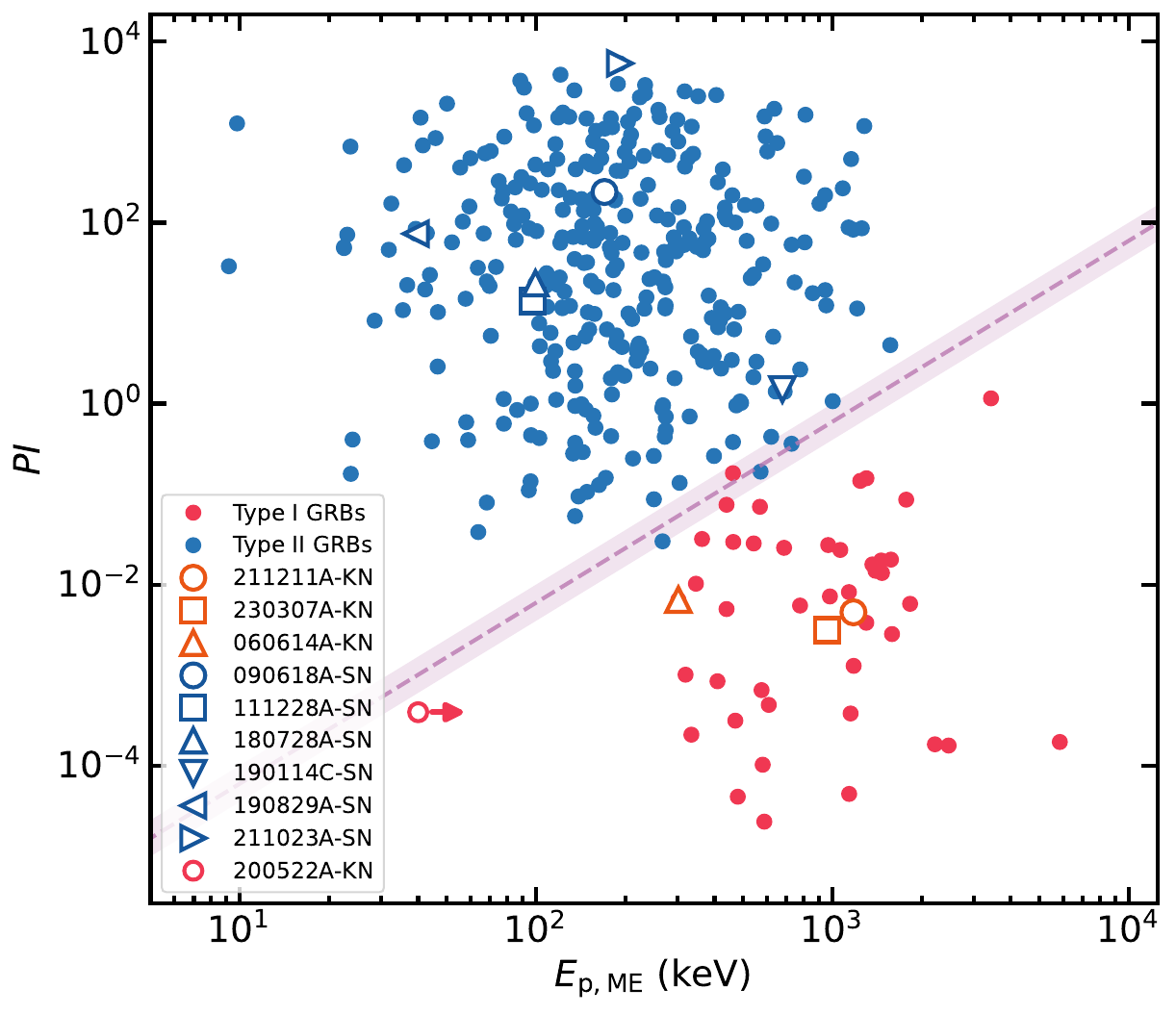}
	\includegraphics[angle=0,scale=0.39]{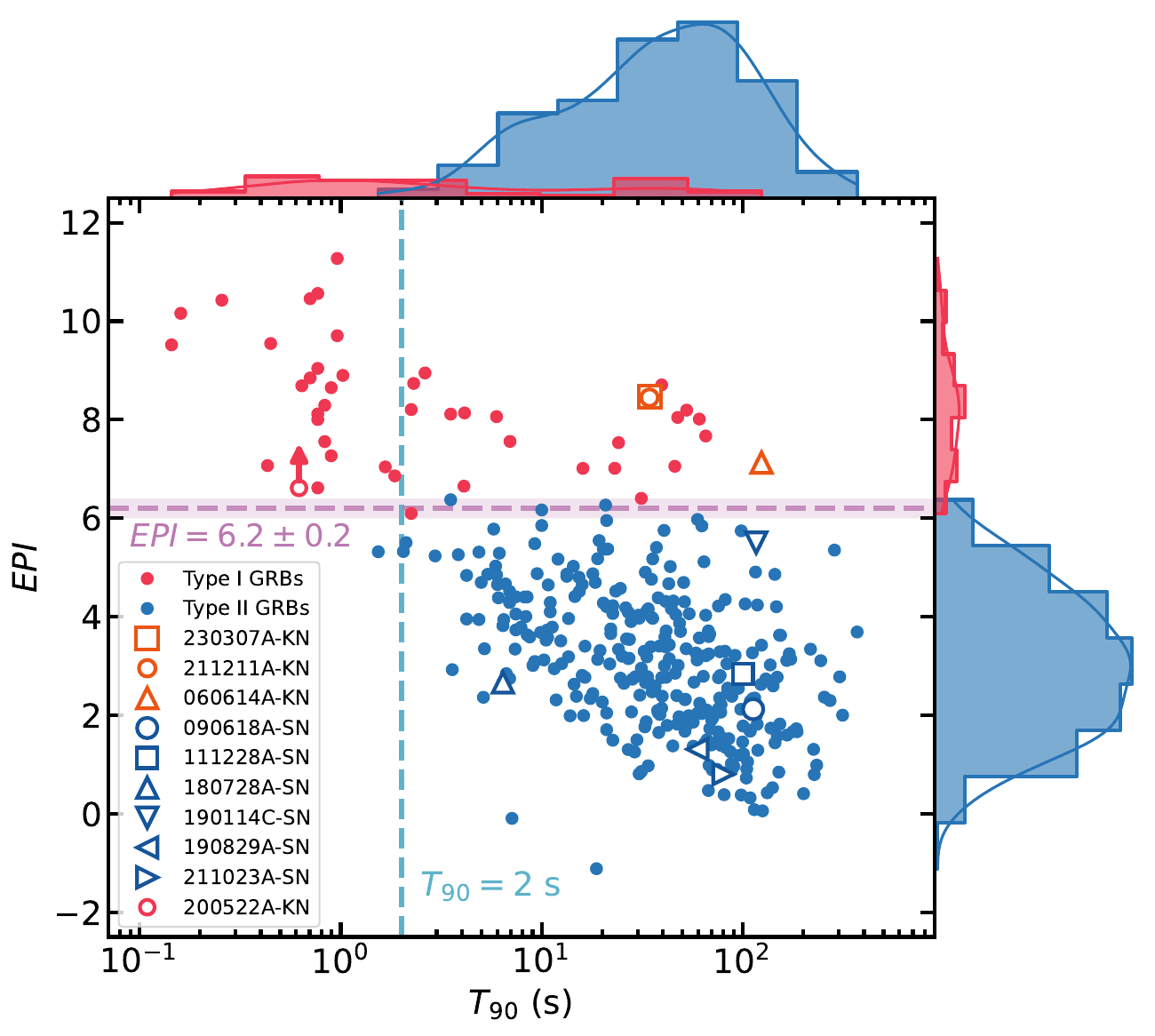}
	\caption{The $PI$--$E_{\rm p,ME}$ plane (left) and the $EPI$--$T_{90}$ plane (right). The hollow red circles represent GRB~200522A with a lower limit of $E_{\rm p}$ at 40 keV. The purple dashed line $PI = 10^{-6.2}E_{p,ME}^{2}$ serves as the dividing line corresponding to $EPI = 6.2$. The shallow purple shaded region represents the range corresponding to the $EPI = 6.2 \pm 0.2$. In the right panel, the histograms along the top and right axes show the distributions of $T_{90}$ and $EPI$, respectively. The light blue dashed line is $T_{90} = 2$ s.}
	\label{figure:par-Ep}
\end{figure*}

Spectral hardness, quantified by $E_{\rm p,ME}$, provides an additional discriminator. 
As shown in Figure~\ref{figure:par-Ep}, Type~I and Type~II GRBs indeed occupy clearly distinct regions on the $PI$--$E_{\rm p,ME}$ plane, and the empirical relation $PI = 10^{-6.2}E_{p,ME}^{2}$ effectively separates the two types of GRBs.
This motivates us to further define a dimensionless $E_{p,ME}$-precursor index,
\begin{equation}
    EPI = {\rm log_{10}} \left[ \frac{(E_{\rm p,ME}/1\ \rm keV)^{2}}{PI} \right].
\end{equation}
Type~I GRBs predominantly have $EPI > 6.2$, while Type~II GRBs predominantly have $EPI < 6.2$.

A transition region of $6.0 \leq EPI \leq 6.4$ contains a small fraction of ambiguous cases (three Type~II and one Type~I GRBs), while outside this range no misclassification is found.
Therefore, caution is needed when classifying GRBs with $EPI$ values between $6.0$ and $6.4$ using $EPI$ alone.


The above results indicate that the distinction among Type~I-S, Type~I-L, and Type~II GRBs is not resulting from artifacts of methodological or hyperparameter choices, but corresponds rather to traditional, but reflects intrinsic structures in the PE and spectral hardness of the ME.

Figure~\ref{figure:par-Ep} (right panel) depicts the $EPI$--$T_{\rm 90}$ plane, which shows the distribution of $EPI$ in the y-axis, and $T_{\rm 90}$ in the x-axis, of the 366 GRBs.
This clearly demonstrates the better distinguishing power of $EPI$ over $T_{\rm 90}$ in classifying Type-I and Type-II GRBs.

We use the adjusted Rand index (ARI) to quantify the similarity between different classifications. 
The ARI ranges from $-$1 to 1, with 1 indicating perfect agreement. 
The ARIs of the $EPI$ and $T_{90}$ classifications relative to the UMAP classification are 0.95 and 0.58, respectively.
Assuming that the UMAP classification reflects the true origins, it is evident that $EPI$ significantly improves the accuracy of origin identification and provides more reliable results for Type~I-L GRBs.

GRB~200522A is a SGRB with PE observed by Swift, and its excess of near-infrared afterglow indicates that it is possibly associated with a KN \citep{2021ApJ...906..127F,2021MNRAS.502.1279O}.
Therefore, GRB~200522A is a Type~I-S GRB.
We use the standard Swift/BAT tool, \texttt{batbinevt}, to analyze the data for GRB~200522A.
Its 15--350~keV light curve consists of a 0.03 s PE, a 0.14 s ME, and a 0.54 s EE, resulting in $T_{\rm 100,WE} = 1.1$ s, with $T_{\rm 100,QE1} = 0.26$ s and $T_{\rm MVT,PE} = 24$ ms. 
Since its $E_{\rm p,ME}$ cannot be reliably estimated, we adopt a conservative lower limit of $\sim 40$ keV \citep{2020ApJ...902...40Z}, corresponding to $EPI \geq 6.61$. 
As shown in Figure~\ref{figure:par-Ep}, GRB~200522A is safely identified as a TypeI GRB.
GRB~060614 and GRB~200522A serve only as two illustrative examples suggesting that the $EPI$ classification may extend beyond Fermi/GBM to other instruments, although the current lack of additional GRBs-KNe events with identifiable PEs limits further validation.
A larger multi-instrument sample is required to test its robustness.

\section{Discussion} \label{sec:discussions}
Our results indicate that mergers and collapsars produce PEs with distinct properties.
Although PE has been extensively studied, its physical origin remains debated, with some models suggesting a common origin with the ME, while others argue them as independent components.

For Type~I GRBs, PE may originate prior to merger.
The pre-merger models can be broadly divided into two branches: the crustal breaking of neutron star (NS), which includes the resonant shattering flare \citep[RSF;][] {2012PhRvL.108a1102T,2020PhRvD.101h3002S,2022MNRAS.514.5385N} and magnetar super flare \citep[MSF;][] {2022ApJ...939L..25Z} models, and the interaction between magnetospheres \citep{2013PhRvL.111f1105P,2018ApJ...868...19W}.

However, a key limitation of these models is their energy budget: PE typically exceeds $10^{49}$ erg and can reach $10^{53}$ erg in extreme cases, whereas RSF and MSF models generally yield luminosities of $\sim10^{47}$ erg s$^{-1}$ \citep{2013ApJ...777..103T}.
Only when the surface magnetic field reaches $10^{14}$-$10^{15}$ G, corresponding to the typical magnetic field of a magnetar, can sufficient energy be provided, while magnetospheric interaction models with similarly strong fields may produce luminosities of $10^{50}$-$10^{51}$ erg s$^{-1}$ \citep{2023ApJ...954L..29D,2012ApJ...757L...3L,2013PhRvL.111f1105P}.

In the post-merger scenario, PE can arise from thermal photosphere emission during the transition from optically thick to optically thin \citep{2000ApJ...530..292M}, and the shock breakout \citep[SBO;][]{2001ApJ...550..410M,2002MNRAS.331..197R}.
Pure photospheric emission predicts quasi-blackbody spectra, inconsistent with observations, while non-thermal components require dissipation via internal shocks \citep{1994ApJ...432..181M,1994ApJ...430L..93R,1997ApJ...490...92K}. 
The fireball-shock model can provide sufficient energy for both the PE and ME \citep{2002ApJ...578..812M}.
In contrast, the SBO model typically provides only $10^{46}$--$10^{47}$ erg, insufficient for most cases. \citep{2012ApJ...747...88N,2018MNRAS.479..588G,2020ApJ...902L..42W}.

The significantly shorter $T_{\rm MVT,PE}$ in Type~I GRBs suggests a smaller emission radius under comparable Lorentz factors.
In summary, the differences in PE properties can be naturally explained by different physical origins, providing indirect evidence that PE in merger-driven GRBs is produced in the pre-merger phase.

\section{conclusions} \label{sec:conclusions}
In this paper, we comprehensively searched the Fermi Catalog and compiled a sample of GRBs with PE.
Based on 9 parameters, we employed t-SNE and UMAP to GRBs with PE, and both methods clearly reveal three distinct clusters.
Interestingly, these three clusters may correspond to Type~I-S, Type~I-L, and Type~II GRBs.

Type~I-S GRBs exhibit significantly shorter temporal properties than Type~II GRBs. 
Type~I-L GRBs are similar to Type~I-S GRBs, except for longer $T_{\rm 100,ME}$ and $T_{\rm 100,WE}$, indicating that both share a common origin distinct from Type~II GRBs, with Type~I-L forming a long-duration subclass within Type~I GRBs. 
When $T_{\rm 100,WE}$ is ignored, GRBs are classified as Type~I and Type~II GRBs, with Type~I GRBs encompassing both Type~I-S and Type~I-L GRBs. 
Although the distributions of spectral lag and MVT overlap between Type~I and Type~II GRBs, shorter values of these parameters remain characteristic of Type~I GRBs.

We find that the PEs of Type~I-S and Type~I-L GRBs differ from those of Type~II GRBs, indicating that PE serves as a key feature to distinguish GRB origins.
These differences likely arise from distinct physical processes, providing direct evidence for the pre-merger origin of the PE in merger-driven GRBs.
We propose a parameter $EPI$ to distinguishing Type~I and Type~II GRBs with PE and adopt a threshold value of $6.2 \pm 0.2$.
If a GRB has an $EPI$ between 6.0 and 6.4, misclassification may occur.
Outside this range, the false positive rate is \emph{zero}, i.e., a GRB with $EPI > 6.4$ can safely be classified as a Type~I GRB, and a GRB with $EPI < 6.0$ can safely be regarded as a Type~II GRB.
The validation using Swift GRB~060614 and GRB~200522A provides two illustrative examples suggesting that the method may also be applicable to GRBs observed by instruments other than Fermi, although further tests with larger cross-instrument samples will be necessary.

The Space-based multi-band astronomical Variable Objects Monitor \citep[SVOM;][]{2016arXiv161006892W} satellite and Einstein Probe \citep[EP;][]{2025SCPMA..6839501Y} will allow for faster and more precise detection of GRBs, enabling further confirmation of special GRB samples through follow-up multi-wavelength observations.
The classification we proposed can quickly provide strong evidence pointing to its origin, thereby enabling a rapid assessment of its observational value.

\begin{acknowledgments}
We acknowledge the use of public data and software provided by the Fermi Science Support Center and the UK Swift Science Data Centre at the University of Leicester. 
This work was supported by the National Natural Science Foundation of China (No. 12273122), National Astronomical Data Center, the Greater Bay Area (No. 2024B1212080003), and the science research grants from the China Manned Space Project (No. CMS-CSST-2025-A13).
F.-W.Z. acknowledges the support from the National Natural Science Foundation of China (No. 12463008). 
\end{acknowledgments}

\facilities{Fermi (GBM), Swift (BAT)}
\software{Astropy \citep{2022ApJ...935..167A}, HEASOFT, GBM Data Tools, scikit-learn \citep{scikit-learn}, umap-learn \citep{mcinnes2018umap-software}}

\clearpage
\appendix
\restartappendixnumbering

\section{the light curves of GRBs-KNe and GRBs-SNe}

\begin{figure*}[htbp!]
	\centering
	\includegraphics[angle=0,scale=0.30]{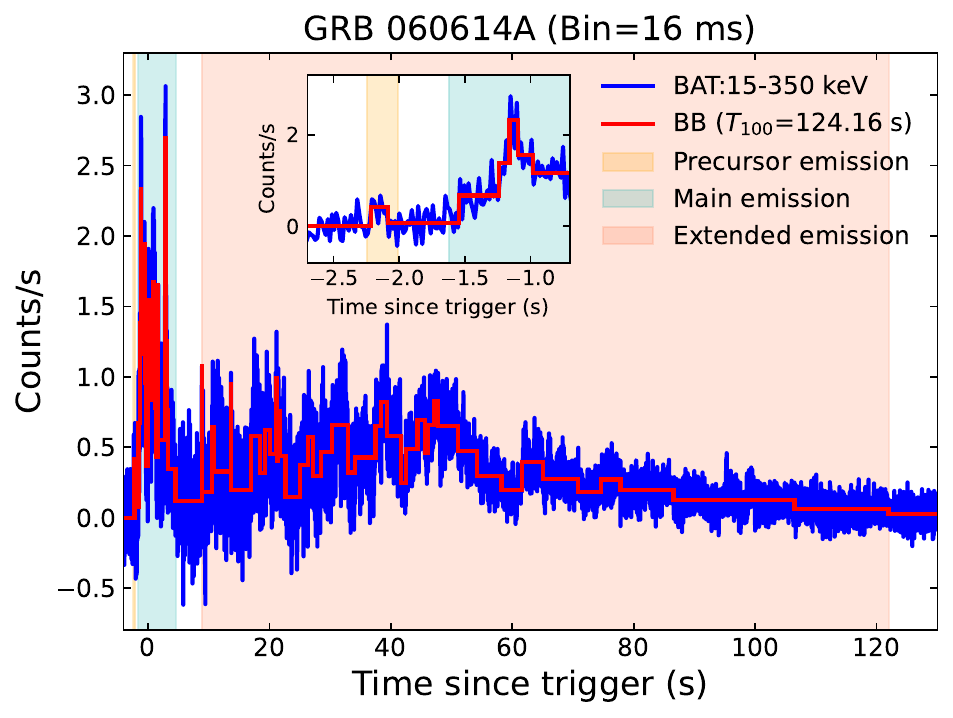}
	\includegraphics[angle=0,scale=0.30]{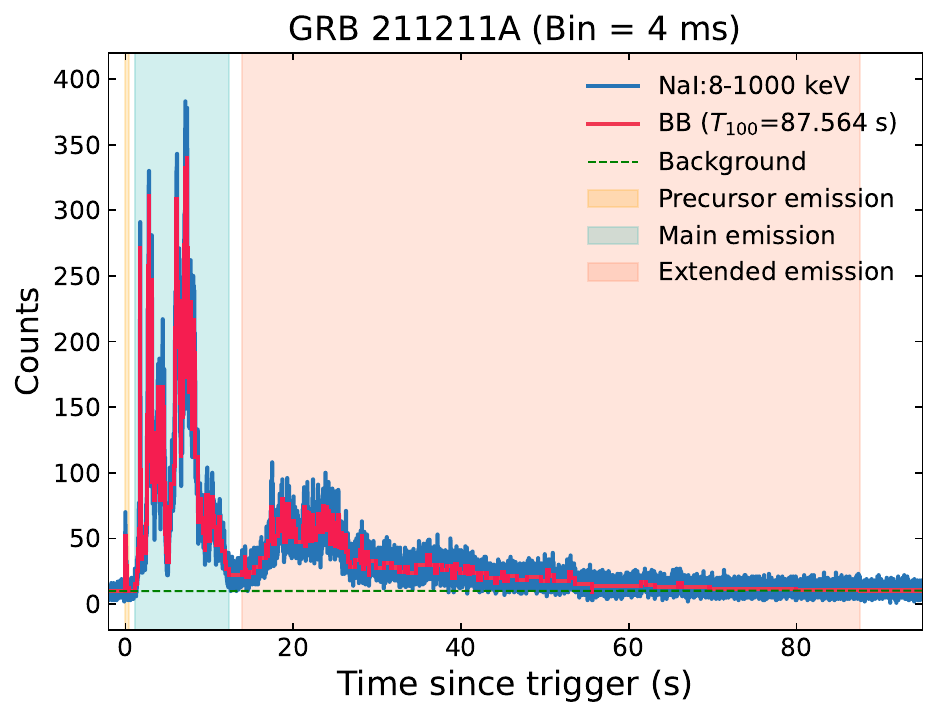}
	\includegraphics[angle=0,scale=0.30]{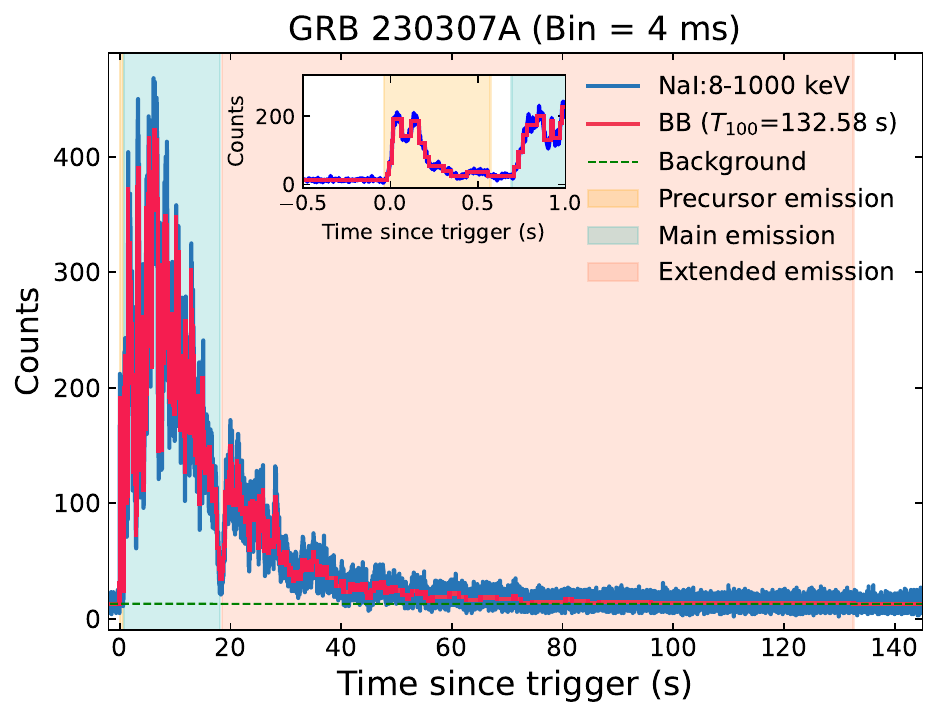}
	\includegraphics[angle=0,scale=0.30]{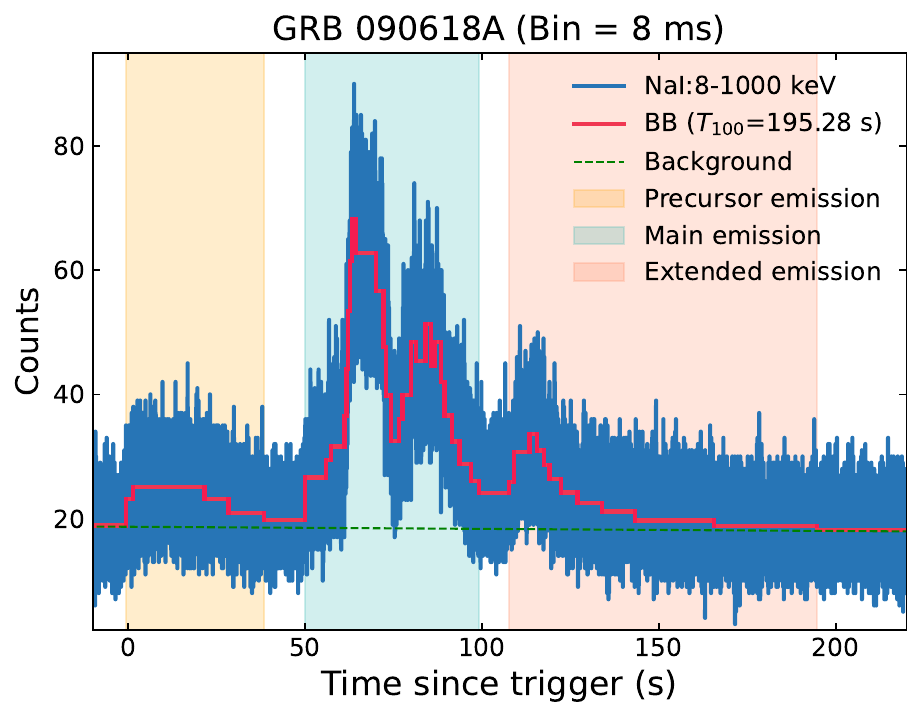}
	\includegraphics[angle=0,scale=0.30]{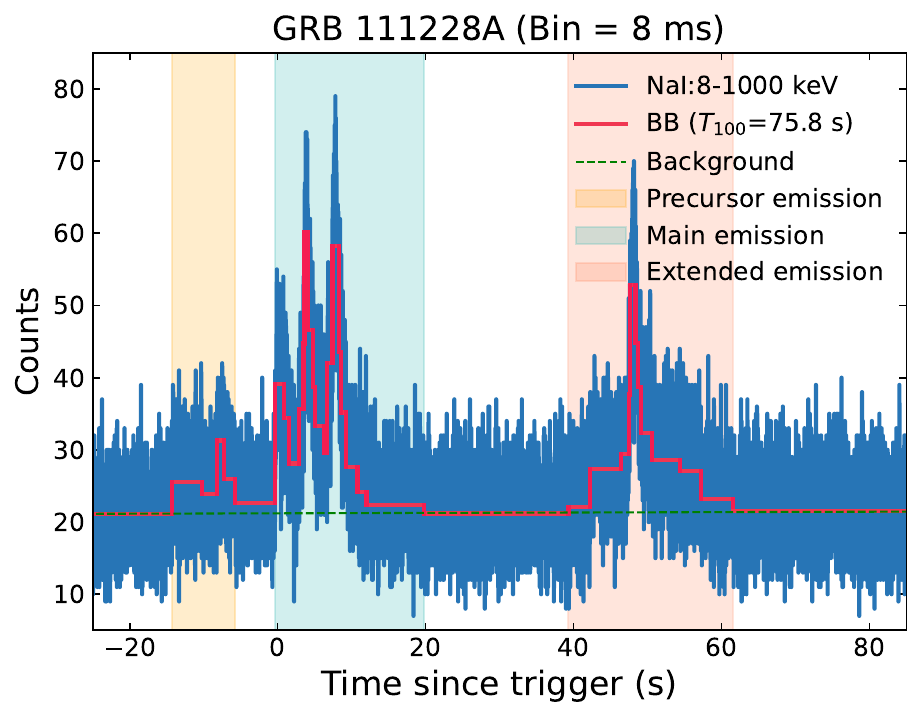}
	\includegraphics[angle=0,scale=0.30]{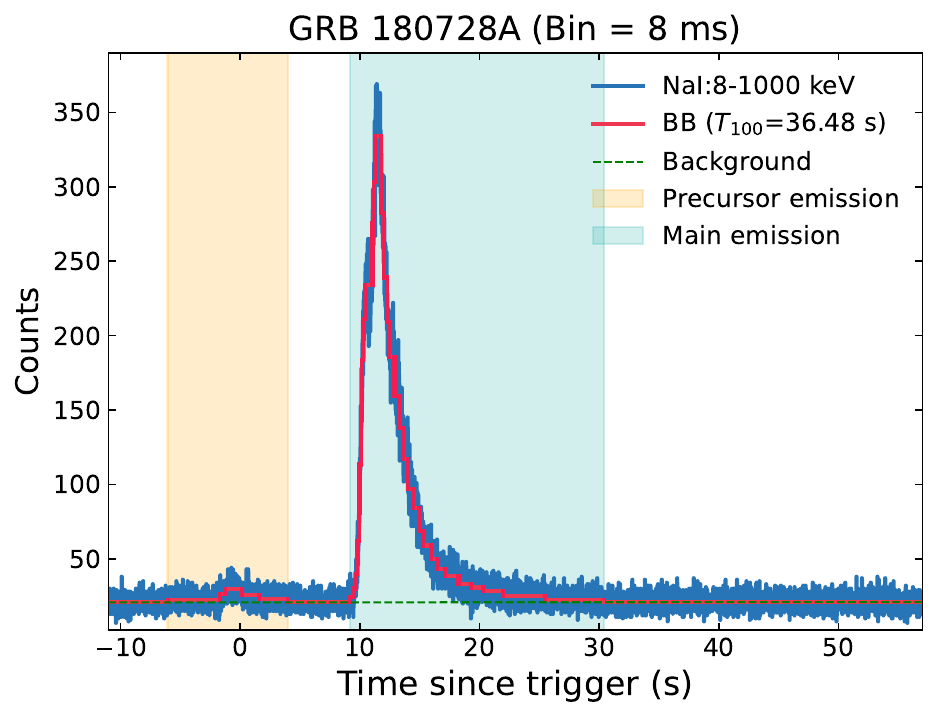}
	\includegraphics[angle=0,scale=0.30]{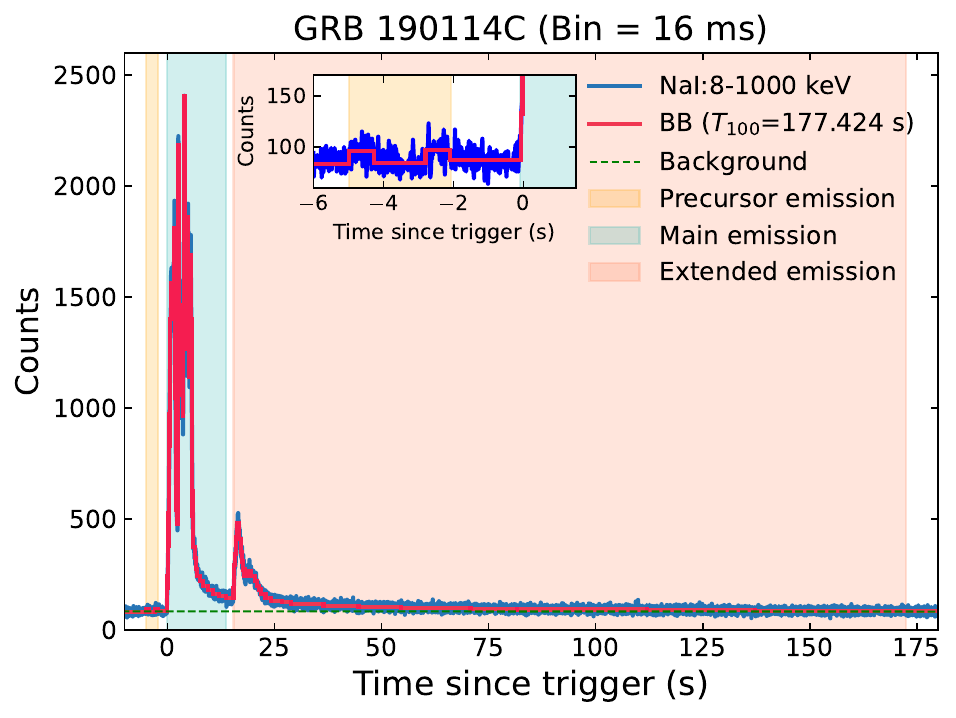}
	\includegraphics[angle=0,scale=0.30]{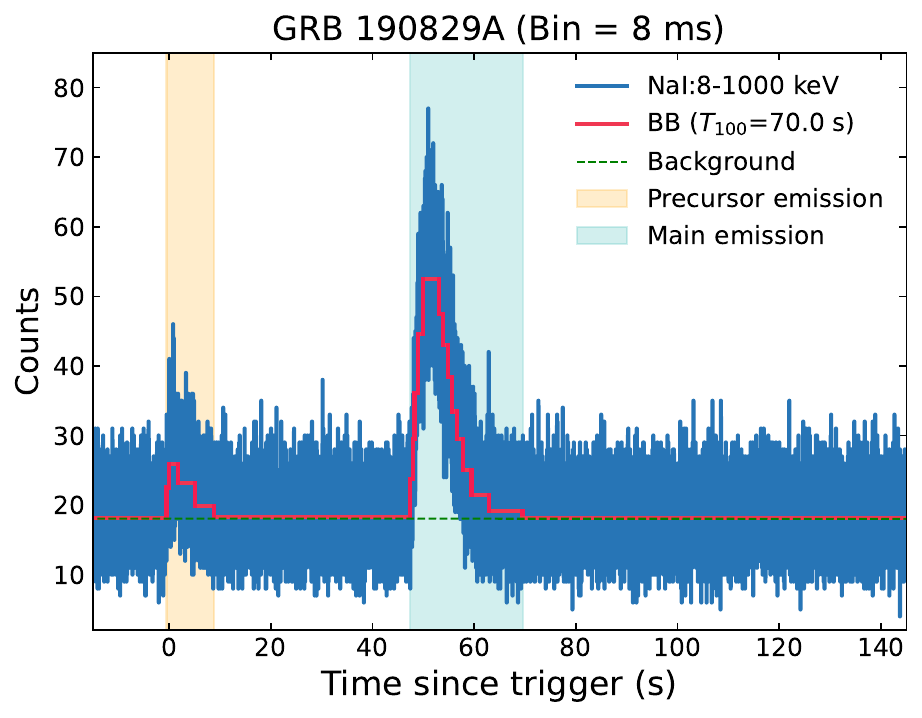}
	\includegraphics[angle=0,scale=0.30]{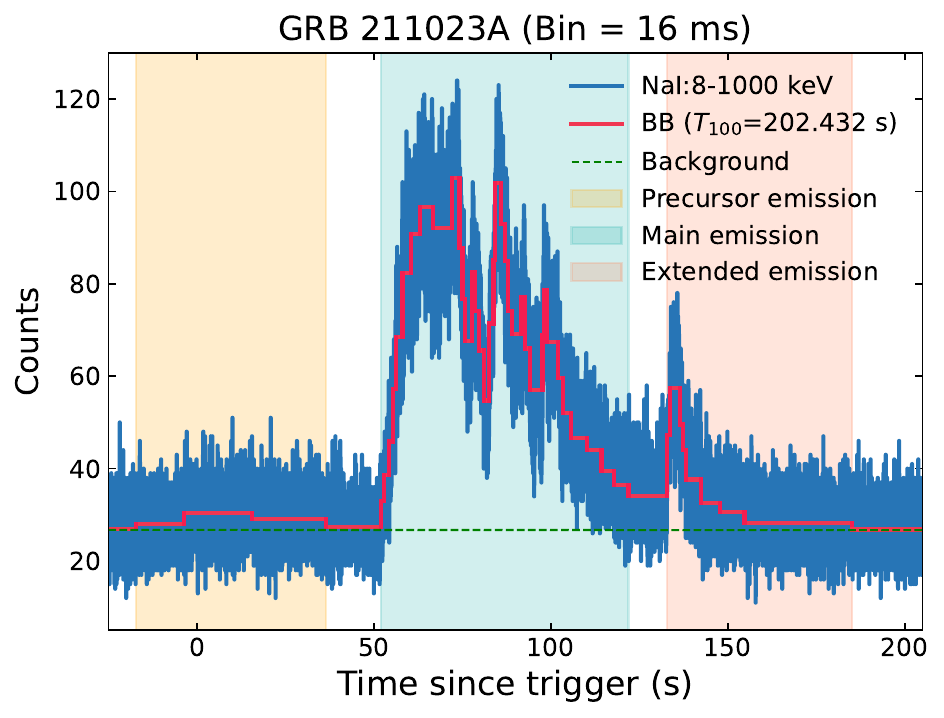}
	\includegraphics[angle=0,scale=0.30]{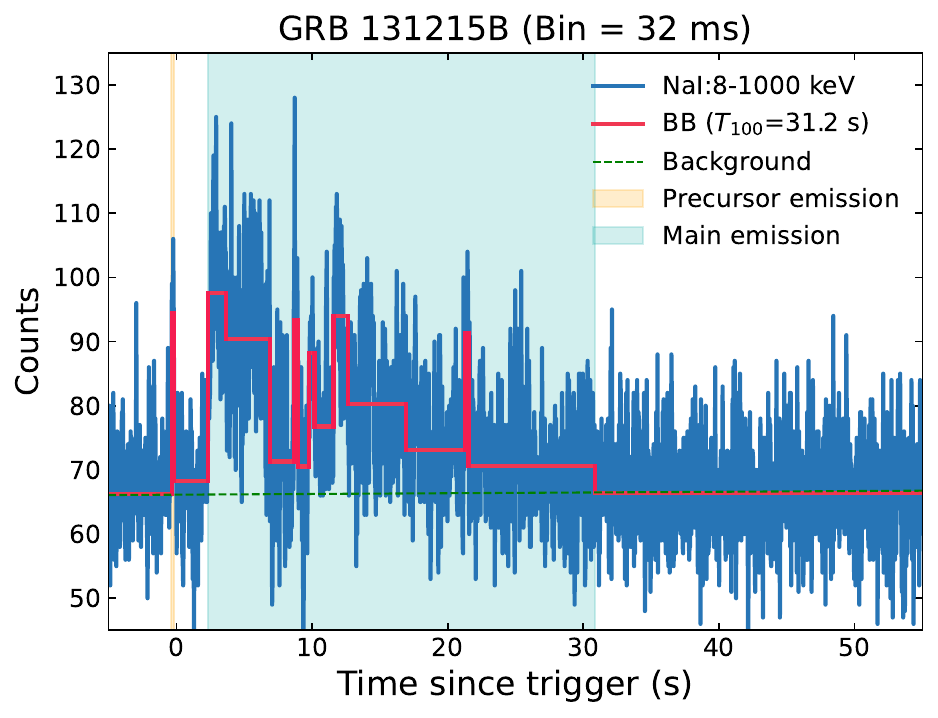}
	\includegraphics[angle=0,scale=0.30]{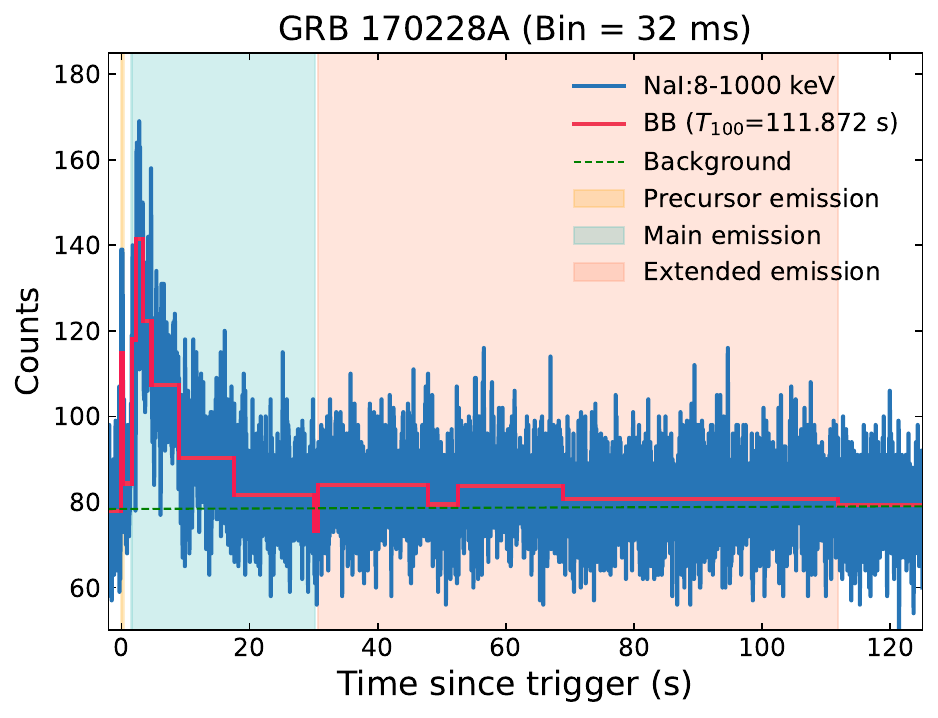}
	\includegraphics[angle=0,scale=0.30]{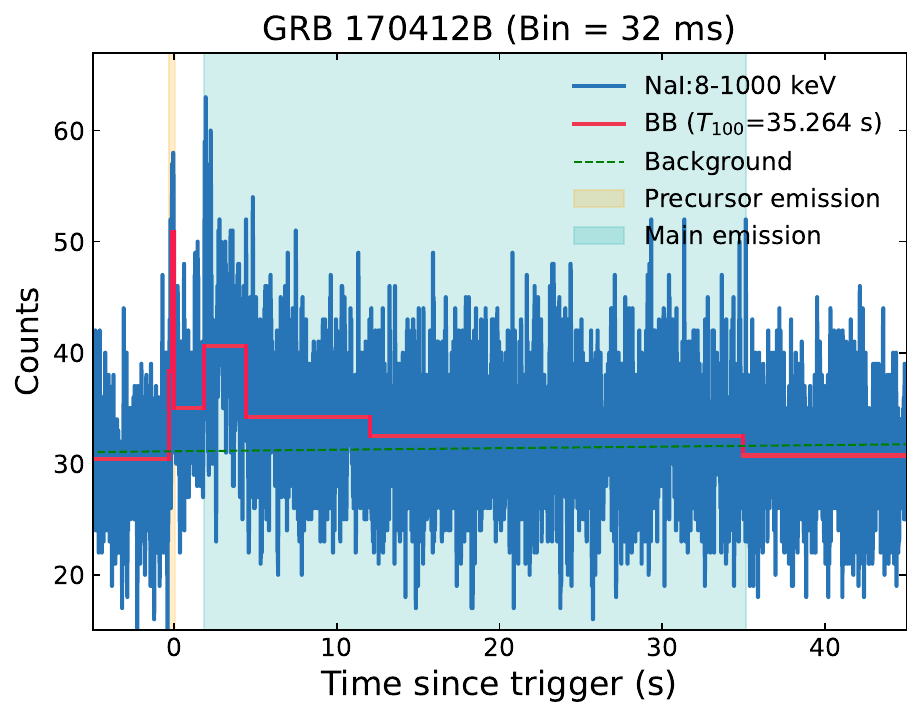}
	\includegraphics[angle=0,scale=0.30]{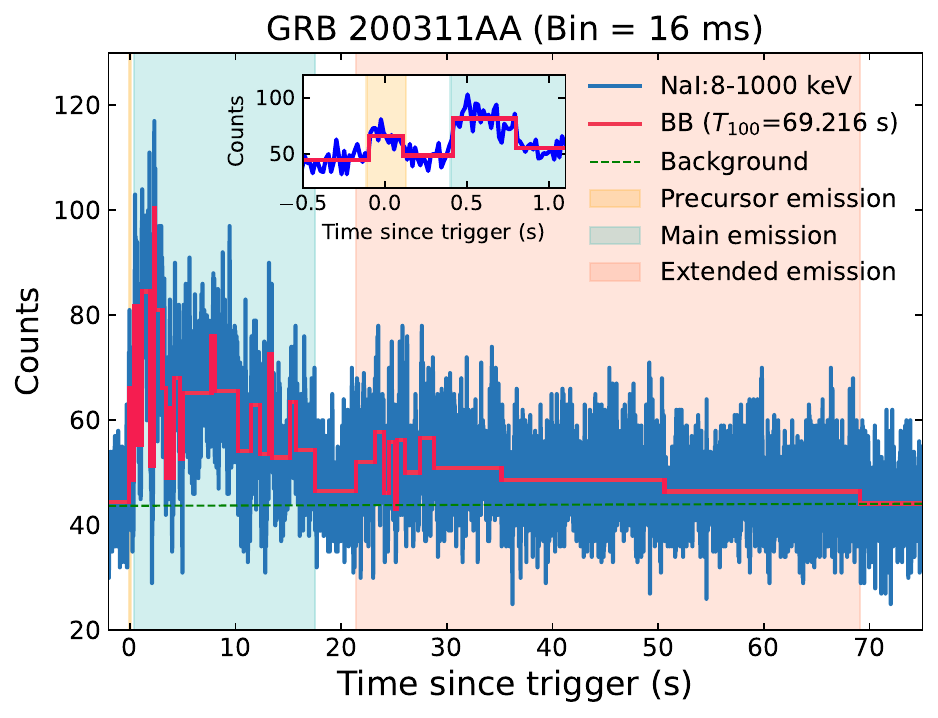}
	\includegraphics[angle=0,scale=0.30]{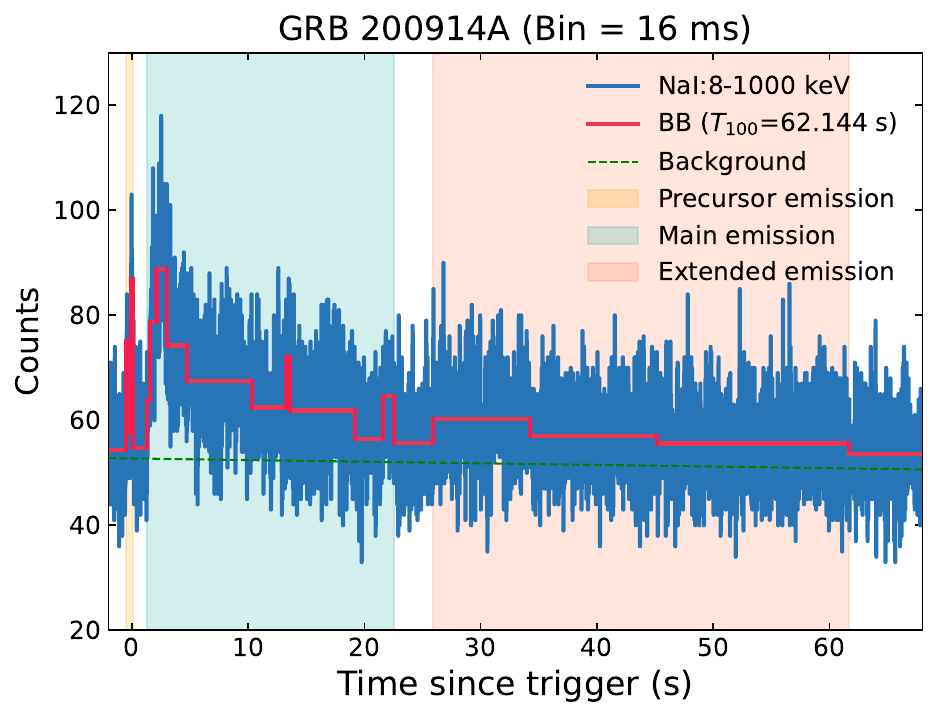}
	\includegraphics[angle=0,scale=0.30]{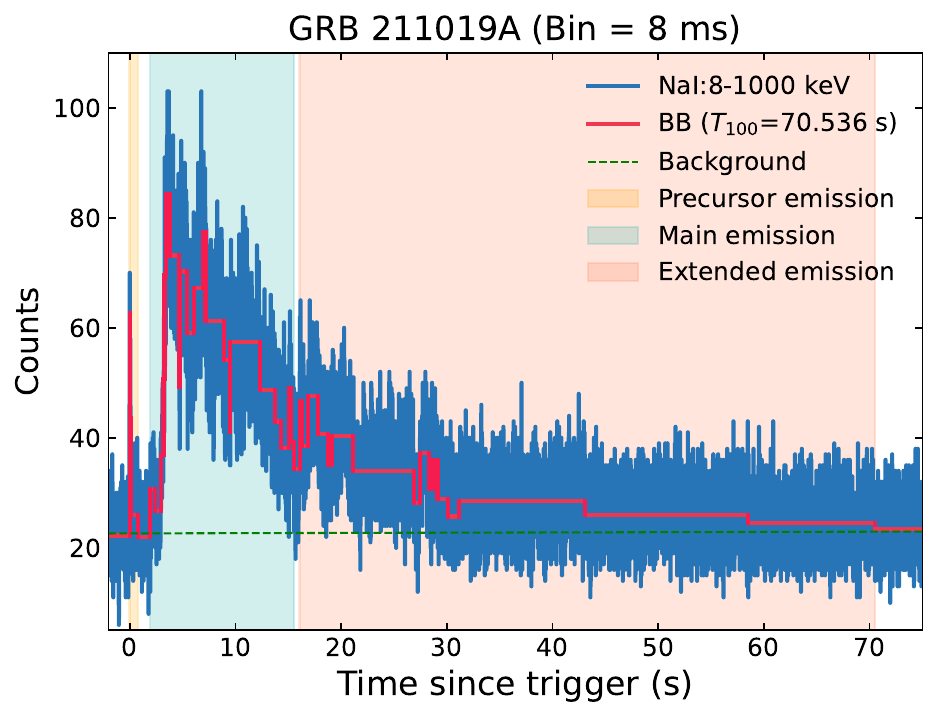}
	\caption{Some example light curves of GRBs. The first row shows GRBs-KNe, the second and third rows show GRBs-SNe, and the fourth and fifth rows show Type I-L GRB candidates. The blue line is the light curve. The red line is the Bayesian blocks. The yellow, green, and red shaded intervals represent PE, ME, EE, respectively.}
	\label{figure:lightcurve}
\end{figure*}

\clearpage
\section{the comprehensive statistical plot}
\begin{figure*}[htbp!]
	\centering
	\includegraphics[angle=0,scale=0.31, trim=0 0 0 150, clip]{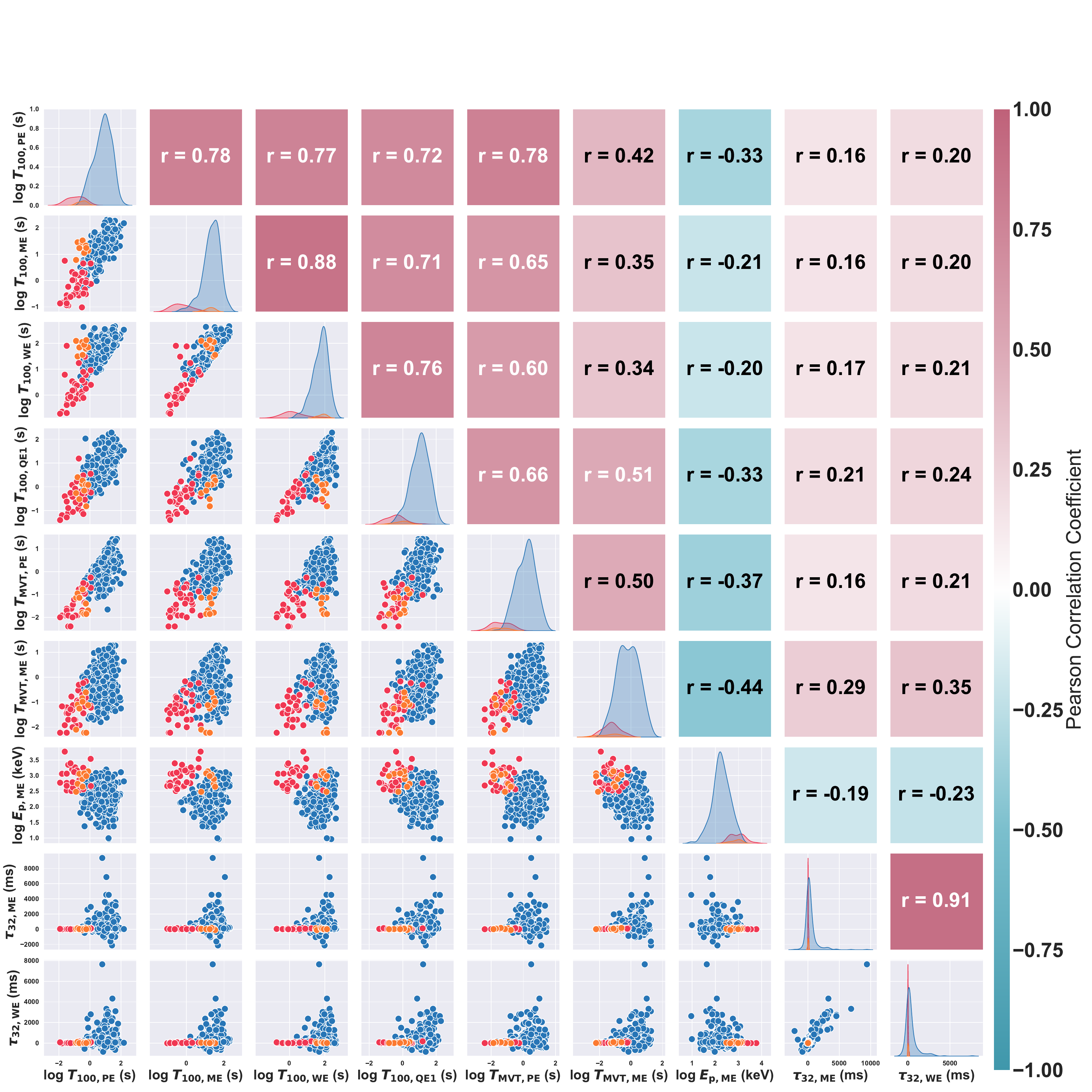}
	\caption{The comprehensive statistical plot of 9 parameters applied for machine learning. The lower-left corner displays scatter plots between each pair of parameters. The diagonal line displays the kernel density plots of the parameters. The red, orange, and blue markers and lines represent Type~I-S, Type~I-L, and Type~II GRBs identified by machine learning, respectively. The upper-right corner displays the Pearson correlation coefficients between the parameters without classification, where red and green colors represent the tight positive and negative correlations, respectively.}
	\label{figure:par}
\end{figure*}

The comprehensive statistical plot of the 9 parameters ($T_{\rm 100,PE}$, $T_{\rm 100,ME}$, $T_{\rm 100,WE}$, $T_{\rm 100,QE1}$, $T_{\rm MVT,PE}$, $T_{\rm MVT,ME}$, $\tau_{\rm 32,ME}$, $\tau_{\rm 32,WE}$, and $E_{\rm p,ME}$) used by machine learning, as well as the Pearson correlation coefficient, are shown in Figure~\ref{figure:par}. 

\clearpage
\section{Empirical Classification of Type I-S and Type I-L GRBs}\label{section:sl}
\begin{figure*}[htbp!]
	\centering
	\includegraphics[angle=0,scale=0.5, trim=0 0 0 0, clip]{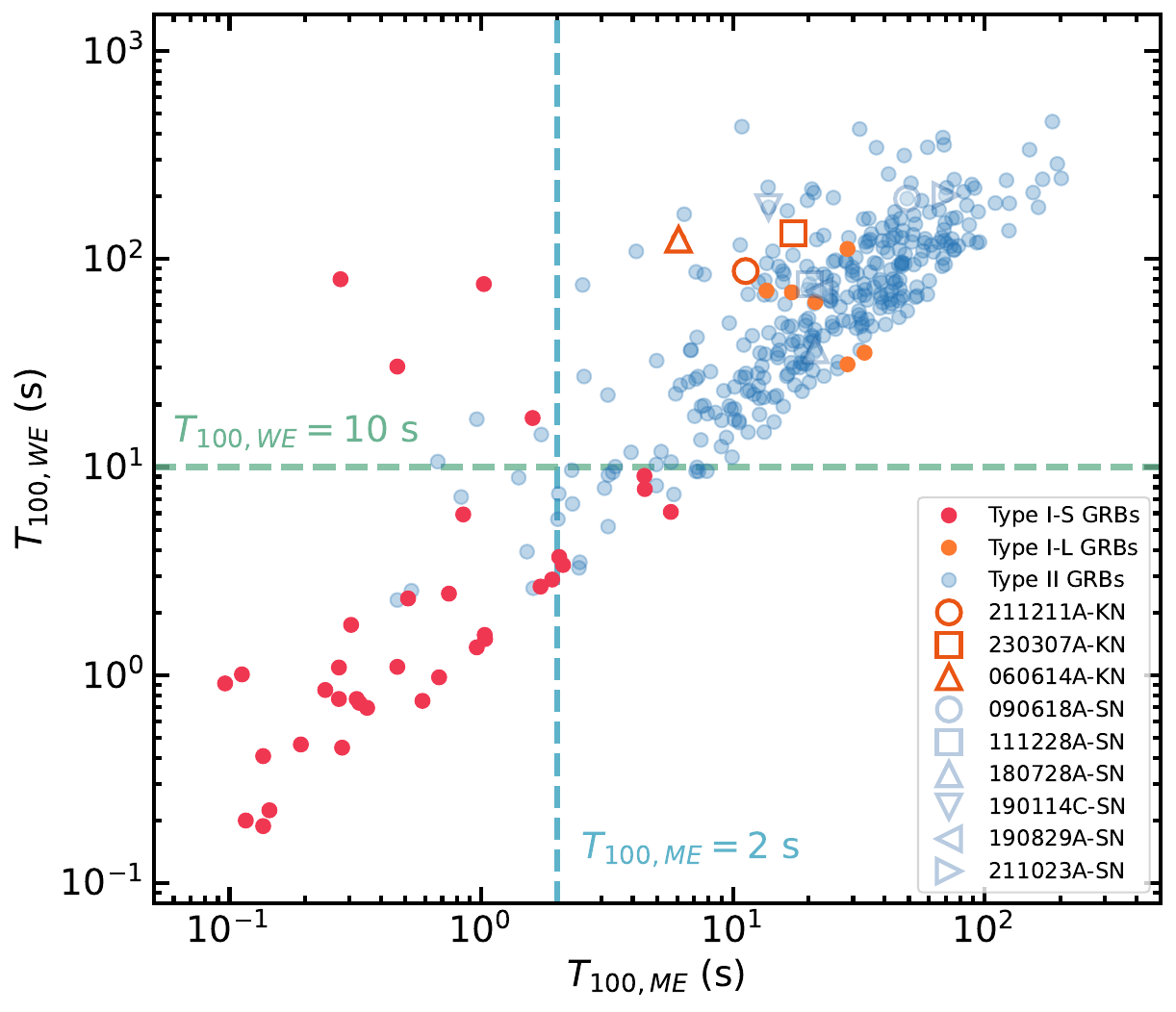}
	\caption{The $T_{\rm 100,WE}$--$T_{\rm 100,ME}$ plane. The light blue and light green dashed lines are $T_{\rm 100,ME} = 2$ s and $T_{\rm 100,WE} = 10$ s, respectively.}
	\label{figure:SL}
\end{figure*}

We note that Type I-S GRBs exhibit systematically shorter $T_{\rm 100,ME}$ and $T_{\rm 100,WE}$ than Type I-L GRBs.
However, as shown in Figure~\ref{figure:SL}, these two subclasses cannot be directly distinguished by a single parameter.
We find that all Type I-L GRBs satisfy $T_{\rm 100,ME} > 2$ s and $T_{\rm 100,WE} > 10$ s, and we therefore empirically classify Type I GRBs with $T_{\rm 100,ME} > 2$ s and $T_{\rm 100,WE} > 10$ s as Type I-L GRBs, while the remaining events are classified as Type I-S GRBs.

\clearpage
\bibliography{ref}{}
\bibliographystyle{aasjournal}


\end{document}